\documentclass[preprint, superscriptaddress]{revtex4}

\usepackage{ifpdf}

    \ifpdf        \usepackage[pdftex]{graphicx}
        \usepackage{epstopdf}
  \else\usepackage{graphicx}\fi
\usepackage[T1]{fontenc}
\usepackage{amssymb}
\usepackage{subfigure}

\begin{document}
\title{Impedance model for the polarization-dependent optical absorption of superconducting single-photon detectors}
\author{E.~F.~C. Driessen}\email{driessen@molphys.leidenuniv.nl}
\author{F.~R. Braakman}
\affiliation{Huygens Laboratory, Leiden University, P.O. Box 9504, 2300 RA Leiden, The Netherlands}
\author{E.~M. Reiger}
\author{S.~N. Dorenbos}
\author{V. Zwiller}
\affiliation{Kavli Institute for Nanoscience, Delft University of Technology, Lorentzweg 1, 2628 CE Delft, The Netherlands}
\author{M.~J.~A. de~Dood}
\affiliation{Huygens Laboratory, Leiden University, P.O. Box 9504, 2300 RA Leiden, The Netherlands}

\begin{abstract}
We measured the single-photon detection efficiency of NbN superconducting single
photon detectors as a function of the polarization state of the incident
light for different wavelengths in the range from 488~nm to 1550~nm.  The polarization contrast varies from $\sim$5\% at 488~nm
to $\sim$30\% at 1550 nm, in good agreement with numerical calculations. We use an optical-impedance model to describe the absorption for polarization
parallel to the wires of the detector. For lossy NbN films, the absorption can be kept constant by keeping the product of layer thickness and filling factor constant. As a consequence, we find that the maximum possible absorption is independent of filling factor. By
illuminating the detector through the substrate, an absorption efficiency of $\sim70\%$ can be reached for a detector on Si or GaAs, without the need for an optical cavity.
\end{abstract}

 \maketitle

\section{Introduction}

Superconducting single-photon detectors
(SSPDs)~\cite{Goltsman:2001p1481}, that consist of a meandering NbN
wire, are an interesting new class of detectors that may outperform
single-photon counting avalanche photodiodes. SSPDs feature a
relatively high quantum efficiency at infrared wavelengths, combined
with low time jitter, low dark counts, and high counting
rates~\cite{Goltsman:2007p272}. This makes these detectors promising
for quantum optical studies and long-distance quantum cryptography
applications~\cite{Takesue:2007p2888}.

A lot of attention has been given to the electronic operation of these
detectors~\cite{Semenov:2003p3080,Kerman:2006p2859,Ejrnaes:2007p2914},
leaving the optical design of the detectors less explored. In fact,
due to the highly anisotropic nature of the wires, the detection
efficiency shows a strong polarization
dependence~\cite{Anant:2008p7675}. This is important, since a common
way to encode quantum information is to use the polarization state
of the photons~\cite{bouwmeester}. Detection of a photon thus
comprises a simultaneous measurement of the polarization, which may
be undesirable for some applications. At the same time, knowledge of
the polarization dependence may simplify experimental schemes that
require a polarization measurement, or can be used to optimize the
detection efficiency.

The efficiency $\eta$ to detect a single photon can be decomposed in an electronic and an optical contribution and can be expressed as
\begin{equation}
\eta = \eta_\mathrm{e} A,
\end{equation}
where $A$ is the optical absorption efficiency of the detector, and $\eta_\mathrm{e}$ is the electronic efficiency of the detector, i.e. the probability that an absorbed photon leads to a measurable voltage pulse across the detector.

The microscopic working principle of the detectors, which is essential to understand $\eta_\mathrm{e}$, is still under active investigation~\cite{Engel:2006p2873,Bell:2007p281}. On a macroscopic level, a photon that is absorbed by the  superconducting wire triggers a temporary loss of superconductivity, which gives rise to a finite voltage pulse across the detector.
The optical absorption efficiency $A$ is determined by the geometry of the detector and the dielectric constants of the substrate and the NbN layer. Since the energy of the incident photons is much larger than the
superconducting gap of the NbN, the complex dielectric constant of the NbN layer at room temperature can be used.

The polarization dependence of NbN SSPDs has been investigated at a single wavelength and compared to finite-difference time domain calculations~\cite{Anant:2008p7675}. In section~\ref{sec:polarization}, we experimentally investigate the wavelength dependence of the polarization contrast, in the range between 488 and 1550~nm, and report a strong dependence of the polarization contrast on the wavelength. We introduce an analytical optical impedance model in section~\ref{Sec:model} to describe the optical absorption in thin lossy films and describe different ways to increase the detection efficiency by changing the parameters of the detector. We find that the optimum thickness is a strong function of the fill fraction, while the maximum achievable absorption is independent of the NbN fill fraction.

The optical impedance model also provides more insight into the cavity enhancement reported for a NbN detector inside a Fabry-Perot type cavity~\cite{Rosfjord:2006p2886,Anant:2008p7675}. We show, in section~\ref{sec:substrate}, that the absorption of the detector is enhanced by a factor $n$, with $n$ the refractive index of the substrate, when the detector is illuminated from the substrate. This factor was not accounted for in earlier work and thus leads to an overestimate of the resonant enhancement. For a high index Si or GaAs substrate this factor becomes dominant and an absorption efficiencie of $\sim 70\%$ can be reached without the need of an optical cavity.


\section{Experimental setup}
\begin{figure}[ht]
\centering %
\includegraphics{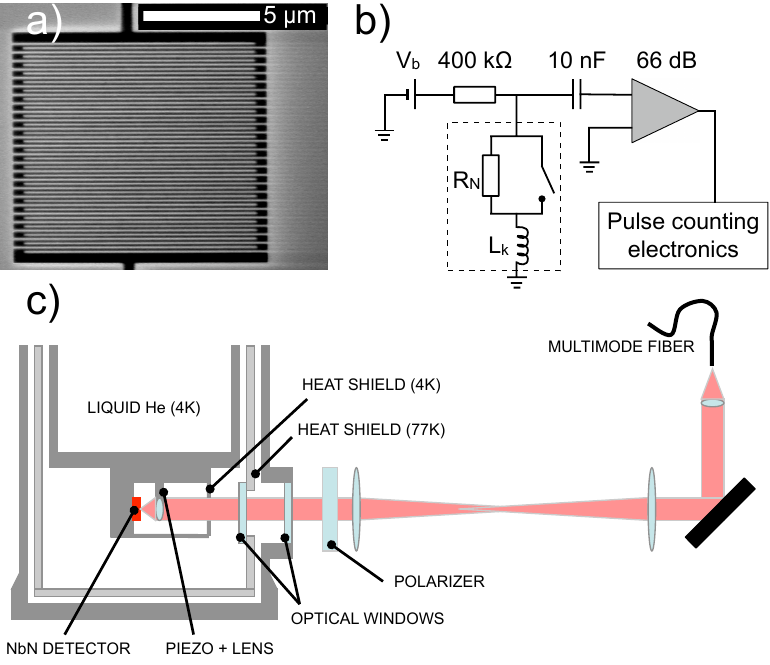}
\caption{(a) SEM image of a NbN SSPD similar to the one used in the experiments. The 100~nm wide NbN line is folded into a meander with an area of 10$\times$10~$\mu\mathrm{m}^2$. (b) Schematic diagram of the readout electronics. The bias current is provided by a voltage source and a resistor of 400~k$\Omega$. The dashed box contains a phenomenologically equivalent circuit of the detector. (c) Schematic overview of the optical setup. Wavelength-filtered light from a lamp is sent through an optical fiber with a 50~$\mu$m core, and is imaged onto the detector through a telescope and a moveable lens mounted inside the cryostat.}\label{fig:setup}
\end{figure}

In our experiments, we used a commercial NbN SSPD~\cite{Goltsman:2007p272}, with an area of 10$\times$10~$\mu$m$^2$. The detector consists of a $\sim$4~nm thick NbN meander on a R-plane sapphire substrate. It has a nominal line width of 100~nm and a filling factor of $\sim 55\%$. Fig.~\ref{fig:setup}(a) shows a scanning electron microscope (SEM) image of a detector similar to the one used in the measurements.

We mounted the SSPD in a $^4$He-cryostat and cooled it to a temperature of $\sim$5~K. The temperature remained constant within 10~mK during each measurement run. Figure~\ref{fig:setup}(b) shows a schematic overview of the electronic circuit used to operate the detector. The detector was biased at 90\% of the critical current through a bias-T with a 400~k$\Omega$ resistor. The equivalent circuit of the detector (dashed box) contains a switch that is closed in the superconducting state. When a photon is absorbed, the switch opens temporarily~\cite{Kerman:2006p2859}. The resulting voltage pulse across the detector is amplified (66~dB) and detected by pulse counting electronics.

Unpolarized light from an incandescent tungsten lamp was wavelength-filtered and sent through a 50 $\mu$m core size multimode optical fiber. The output of the fiber was imaged onto the detector using a telescope and a lens mounted on a piezo stage inside the cryostat, as shown in Fig.~\ref{fig:setup}(c). To probe the polarization dependence of the detection efficiency, a linear polarizer with an extinction ratio better than $100:1$ for the wavelength range of interest was placed in the parallel part of the beam. To probe the wavelength dependence, we used different narrow bandpass filters ($\leq$10~nm FWHM) in combination with several edge filters to ensure that the light on the detector was monochromatic.

\section{Polarization dependence}\label{sec:polarization}

\begin{figure}[ht]
\centering %
\includegraphics[width=.5\linewidth]{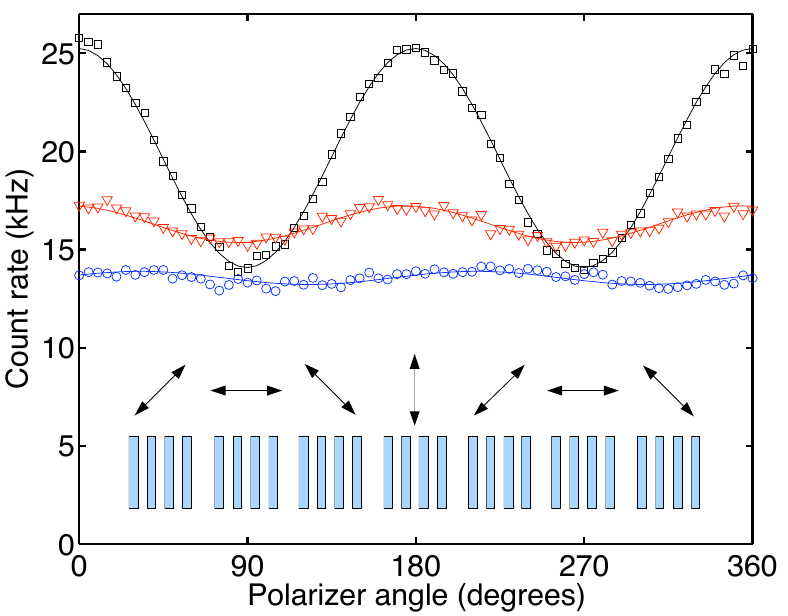}
\caption{Count rate of the SSPD (corrected for stray light counts) as a function of linear polarization, for a wavelength of 1550~nm (black squares) and 532~nm (red triangles). The blue circles show the count rates for 1550~nm light when two wedge depolarizers (under a relative angle of 45$^\circ$) are placed after the polarizer. The insets show the orientation of the $E$-field relative to the detector, for the different polarizer settings. The solid curves are sinusoidal fits to the data, used to extract the polarization contrast.}\label{fig:counts}
\end{figure}

Figure~\ref{fig:counts} shows the count rate of the detector as a function of linear polarization for a wavelength of 1550~nm (black squares) and 532~nm (red triangles). Note that the absolute count rates at different wavelengths cannot be compared directly, due to a difference in incident power. The insets show the orientation of the $E$-field relative to the detector. The measured count rates follow a sinusoidal dependence as a function of polarization and are minimal when the $E$-field is perpendicular to the lines of the detector.

We define the polarization contrast $C$ as
\begin{equation}
C = \frac{N_\parallel -N_\perp}{N_\parallel + N_\perp},
\end{equation}
where $N_\parallel$ and $N_\perp$ are the count rates of the
detector when the light is polarized parallel and perpendicular to
the wires, respectively. This definition of the contrast is a direct
measure for the polarization effects, independent of the electronic
quantum efficiency ($\eta_\mathrm{e}$), and the incident power. We
extract the contrast from the sinusoidal fits to the data (solid
curves in Fig.~\ref{fig:counts}). It varies with the wavelength of
the incident light and is independent of the bias current and
temperature of the detector in our experiment.

The blue circles in Fig.~\ref{fig:counts} show the count rate as a
function of polarizer angle, at a wavelength of 1550~nm, when two
wedge depolarizers under a relative angle of 45$^\circ$ were placed
after the polarizer. These wedge depolarizers effectively depolarize
the incident light by imposing a position-dependent
rotation of the polarization. Indeed, the polarization contrast in this
case is reduced to below $3\%$. The lower average count rate can be
attributed to the extra four air-glass interfaces in the optical
setup, leading to an increased reflection of the incident light.

The polarization effect can be understood by comparing the periodic
structure of the detector to that of a wire grid
polarizer~\cite{Bird:1960p2569} that consists of a grid of parallel,
highly conductive metal wires with a subwavelength spacing. For a
perfect conductor the $E$-field should be perpendicular to the metal
surface. As a consequence, only light with a polarization
perpendicular to the wires is efficiently transmitted. A similar
argument holds for lossy metals, albeit that in this case the field
penetrates into the metal, leading to absorption. This absorption is
largest when the $E$-field is parallel to the wires, since in this
case the field penetrates more into the metal.


For the typical dimensions and spacing of the NbN wires, an
effective medium approach that is accurate for both polarizations is
difficult~\cite{Aspnes:1982p2553,Pitarke:1998p2559}. Instead, we
calculated the absorption at normal incidence for an
infinitely-sized detector, using the rigorous coupled-wave analysis
(RCWA) developed in Ref.~14\nocite{Moharam:1995p2558}. This method finds
an exact solution of Maxwell's equations by expressing the
electromagnetic fields in the different materials as a summation
over all diffraction orders. The Fourier components of the
periodic dielectric constant couple the diffraction orders in the
patterned region. The continuity of the parallel component of the
wavevector, together with the boundary conditions for the $E$ and
$H$ fields fully determine the field in all regions. From this the
intensity in all reflected and transmitted diffraction orders can be
calculated. The absorption in the grating is then simply given by
$A=1-R-T$, where $R$ and $T$ are the reflected and transmitted intensity.

The effects of focusing of the incident beam can be taken into account by decomposing the beam into plane
waves with wave vector $\vec{k}$. Each of these plane waves will
experience a different absorption $A(\vec{k})$. The effect of finite
detector size can be incorporated in a similar way, by multiplying the beam profile in the near field by an aperture function $D(\vec{r})$ which is 1 at the
location of the detector, and 0 elsewhere. Taking both into account,
the total absorption is given by the convolution integral
\begin{equation}
A = \int_{\vec{k}}A(\vec{k})\left[u(\vec{k})\ast
D(\vec{k})\right]^2d\vec{k},\label{eq:fourier}
\end{equation}
where $u(\vec{k})$ is the Fourier transform of the beam profile, and
$D(\vec{k})$ is the Fourier transform of the aperture function
$D(\vec{r})$.

The $k$-spread of the incident waves is determined by the detector
size [determining the spread in $D(\vec{k})$] and the numerical
aperture of the last lens in the illuminating system, determining
the spread in $u(\vec{k})$. The latter is the most important factor
in our experiment, since we used a large-NA lens to focus the
incoming light onto the detector. Calculations of the absorption of
the grating as a function of angle of incidence (i.e., as a function
of $\vec{k}$) show however, that the absorption only varies
appreciably from the absorption at normal incidence for angles of
incidence corresponding to $\mathrm{NA}>0.5$. Therefore, the total
absorption given by Eq.~(\ref{eq:fourier}) can be approximated by a
product of the absorption coefficient at normal incidence and the
total intensity impinging on the (finite-sized) detector. This
justifies the use of a plane wave calculation in the rest of this Paper.

To calculate the absorption efficiency,  we used the nominal
structure parameters of the detector, and tabulated values of the
dielectric constant of the sapphire
substrate ($n_\mathrm{sapphire}=1.74$ at 1550~nm)~\cite{palik3:sapphire}. For the wavelength-dependent
dielectric constant of NbN,  a Drude model~\cite{Tanabe:1988p2544}
was used, giving a refractive index $n_\mathrm{NbN}=5.5+6.3i$ at a
wavelength of 1550~nm. This value is close to the value reported in Ref.~7\nocite{Anant:2008p7675}, for a thicker NbN film.

\begin{figure}[ht]
\centering %
\includegraphics{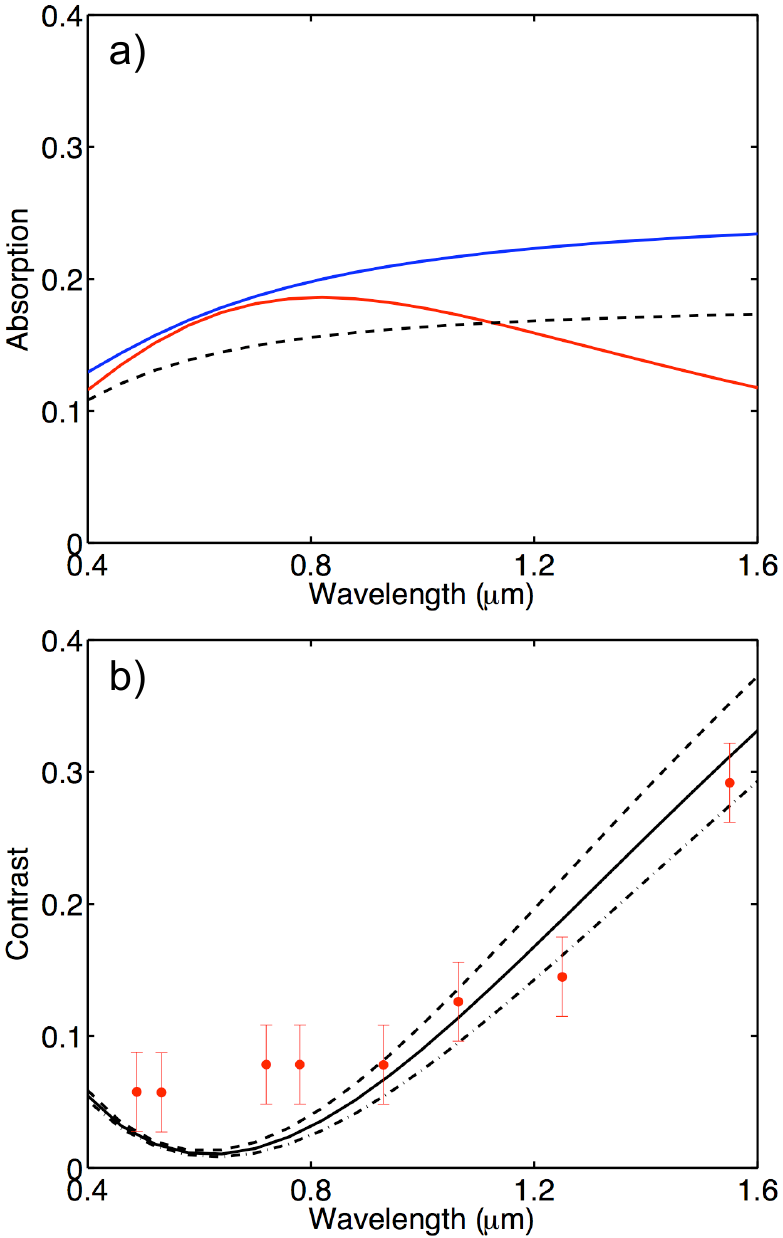}
\caption{(a) Calculated absorption efficiency of a NbN grating as
function of wavelength for polarization parallel (blue curve) and
perpendicular (red curve) to the lines of the detector. For
comparison, the dashed line shows the calculated absorption of an
unpatterned film multiplied by the filling factor of NbN. (b)
Measured (red dots) and calculated (black curves) polarization
contrast as a function of wavelength. The calculations are shown for
a filling factor of 52\% (dashed), 55\% (solid), and 58\%
(dash-dotted) and a film thickness of
4.5~nm.}\label{fig:calculations}
\end{figure}

Figure~\ref{fig:calculations}(a) shows the calculated absorption for
polarization parallel (blue line) and perpendicular (red line) to
the wires, as a function of wavelength. The absorption for
parallel-polarized light monotonously increases with wavelength,
whereas the absorption for perpendicular polarization goes through a
maximum and decreases for wavelengths above $800$~nm. This leads to
a higher polarization contrast for longer wavelengths.

For comparison, the dashed line in Fig.~\ref{fig:calculations}(a) shows
the absorption of an unpatterned film, multiplied by the filling
factor of NbN, as was suggested in Ref.~1\nocite{Goltsman:2001p1481}.
This estimate deviates over the entire wavelength range from the polarization-averaged result obtained by RCWA, which shows that for structures with features smaller than the wavelength of light, a more refined model is needed. We will discuss this refined model in Sec.~\ref{Sec:model}. The fact that the absorption decreases for both the parallel polarization and for the closed film is mostly due to dispersion of the dielectric constant of the NbN material, $\epsilon_\mathrm{NbN}$.


In Fig.~\ref{fig:calculations}(b) we compare the measured
polarization contrast (red dots) to the results of the calculations
(black solid curve), as a function of wavelength. For comparison,
the calculated contrast is shown for filling factors of 52\% (dashed
curve) and 58\% (dash-dotted curve) as well. The experimentally
observed contrast varies between $\sim$5\% and $\sim$30\% and
increases with wavelength. The error bars on the experimental points represent slight variations in the measured polarization contrast during different measurement runs, as well as a slight polarization in the illuminating light source, of $\sim 1\%$. We attribute the fact that the
calculation and the measurements differ for lower wavelengths to the
fact that we used literature values for the dielectric constant
of NbN. It is known that the dielectric constant of NbN varies as a function of the
deposition parameters~\cite{Tanabe:1988p2544} and may depend on the
film thickness as well~\cite{Lee:2008p10411}. Additional calculations (not shown) reveal, that for lower wavelengths, the polarization contrast is increasingly sensitive to small variations in the dielectric constant of NbN.


It has been shown that the linear-polarization dependence can
be removed by changing the design of the detector~\cite{Dorenbos:2008p10421}. A spiraling detector breaks the translational symmetry that causes the strong polarization contrast. The optical absorption in such a detector, however, will be lower than the maximum obtainable for parallel-polarized light, due to the fact that in these detectors, partial screening of the electric field is always possible.

\section{Discussion}\label{Sec:model}
\subsection{An optical impedance model for the absorption of a metal film}\label{sec:impedance}
In order to gain some physical insight into the absorption in the
detector, we start out by describing the absorption of a film of
thickness $d$ with a complex dielectric constant $\epsilon_2$,
embedded between two dielectrics with refractive index $n_{1}$ and
$n_{3}$, respectively. The film is illuminated from the medium with
index $n_1$.

We can define the optical impedance of a medium $i$ with refractive
index $n_i$ as
\begin{equation}
\eta_i = \frac{\eta_0}{n_i},
\end{equation}
where $\eta_0 = \sqrt{\mu_0/\epsilon_0} = 377\Omega$ is the
impedance of the vacuum. The reflection and transmission of the
layered system are given by~\cite{ramo}
\begin{eqnarray}
R &=&
\left|\frac{\eta_\mathrm{load}-\eta_1}{\eta_\mathrm{load}+\eta_1}\right|^2,\\
T &=& \frac{\eta_1}{\eta_3}
\left|\frac{2\eta_\mathrm{load}}{\eta_\mathrm{load}+\eta_1}\right|^2,
\end{eqnarray}
where $\eta_\mathrm{load}$ is the combined load impedance of the
film and the backing substrate. The absorption of the film is again given
by $A=1-R-T$.

If we assume that the film is thin enough to neglect interference
effects ($k_0d \ll 1$),  the load impedance is given by~\cite{Kornelsen:1991p2549}
\begin{equation}
\eta_\mathrm{load} \approx \frac{R_\square\eta_3}{R_\square +
\eta_3},
\end{equation}
where $R_\square \approx \eta_0 / k_0d~\mathrm{Im}~\epsilon_2$ is
the square resistance for a highly absorbing
($\mathrm{Im}~\epsilon_2 \gg \mathrm{Re}~\epsilon_2$) film, and
$k_0$ is the wave vector of the light in vacuo. With these
assumptions, we can write the absorption in the film as
\begin{equation}
A =
\frac{4}{\eta_1R_\square}\left(\frac{\eta_1R_\square\eta_3}{\eta_1+R_\square+\eta_3}\right)^2=
4n_1\frac{k_0d~\mathrm{Im}~\epsilon_2}{\left(n_1+n_3+k_0d~\mathrm{Im}~\epsilon_2\right)^2}.\label{eq:absorption}
\end{equation}
The absorption of the film reaches a maximum value $A_\mathrm{max} =
n_1/(n_1+n_3)$ for a square resistance given by
\begin{equation}
R_\square = \frac{\eta_1\eta_3}{\eta_1+\eta_3}.
\end{equation}
Note that the maximum possible absorption is a function of the refractive indices of the surrounding media only. The optimal value of $R_\square$ to reach this maximum can be obtained by tuning the film thickness $d$.

\subsection{The effect of film thickness}
\begin{figure}[ht]
\centering%
\includegraphics[width=.5\linewidth]{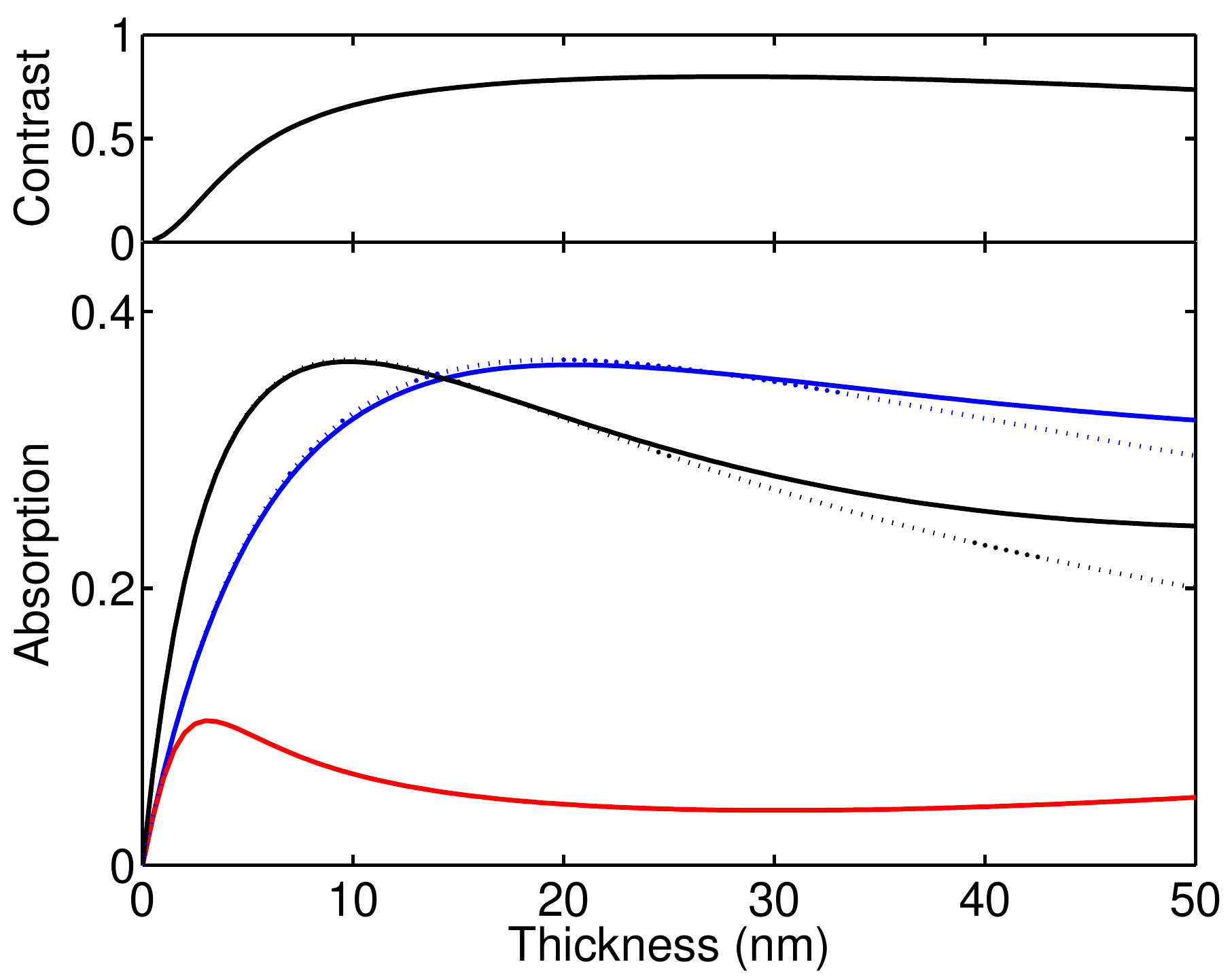}
\caption{Calculated absorption at a wavelength of 1550~nm as a function of film thickness. The black curve gives the absorption for a closed film, the blue and red curves for a detector with lattice period 200~nm and filling factor 0.5, for polarization parallel (blue) and perpendicular (red) to the wires of the detector. The dotted curves are calculated using the impedance model of section~\ref{sec:impedance}, whereas the solid curves are exact calculations using RCWA. The top graph shows the calculated polarization contrast.\label{fig:thickness}}
\end{figure}

Figure~\ref{fig:thickness} shows the absorption and the polarization contrast of a film  of NbN,
embedded between air ($n_1=1$) and sapphire ($n_3 = 1.74$), as a
function of the film thickness. The solid curves show the calculated
absorption using the rigorous coupled-wave analysis described before, while the dotted curves are obtained from
the impedance model.

For a closed film (black curves), there is a distinct maximum of absorption, that
occurs at a thickness
\begin{equation}
d = \frac{n_1+n_3}{k_0~\mathrm{Im}~\epsilon_2}.\label{eq:thickness}
\end{equation}
For thinner films, the transmission through the film is too high to
get maximal absorption, whereas for thicker films, reflection
dominates.

The blue and red curves in Fig.~\ref{fig:thickness} show the
absorption for a detector with filling factor 0.5 and lattice period
200~nm, for polarization parallel and perpendicular to the wires,
respectively. The thickness for which the absorption in the patterned film is maximum, is higher than the optimal thickness for the closed film. The dotted line is
calculated using the impedance model of section~\ref{sec:impedance},
taking an effective dielectric constant for the absorbing film,
given by~\cite{Aspnes:1982p2553}
\begin{equation}
\epsilon_\mathrm{eff} = (1-f)\epsilon_\mathrm{slits}+f\epsilon_\mathrm{NbN},
\end{equation}
where $f$ is the filling factor of the metal, and
$\epsilon_\mathrm{slits}$ is the dielectric constant of the material
in the slits, typically air. Since only the imaginary part of
$\epsilon_\mathrm{eff}$ determines the absorption in the film, the
absorption of the detector can simply be calculated by multiplying
the thickness of the film by the filling factor. For the
polarization perpendicular to the wires of the detector, it is not
so straightforward to define an effective dielectric
constant for the patterned
film~\cite{Aspnes:1982p2553,Pitarke:1998p2559}. For this polarization the light is concentrated in the air
slits and the effective dielectric constant is closer to that of air. Therefore the condition $\mathrm{Im}~\epsilon_2 \gg
\mathrm{Re}~\epsilon_2$ used to define the impedance model, breaks down for this polarization.

Surprisingly, the calculation also shows that it is
easily possible to construct a detector where the absorption for
parallel polarization is larger than the absorption of an
unpatterned film of the same thickness. Since the electronic
efficiency of the detector, $\eta_\mathrm{e}$, strongly depends on
the thickness of the
metal~\cite{Verevkin:2004p10166,Jukna:2008p6859}, it is important to
realize that the absorption for parallel-polarized light is a
function of $df~\mathrm{Im}~\epsilon_2$. A reduction in thickness of
the detector, to increase the electronic efficiency, can thus be
countered by increasing the filling factor accordingly.

\subsection{Illuminating through sub- or superstrate}\label{sec:substrate}
\begin{figure}[ht]
\centering%
\includegraphics[width=.5\linewidth]{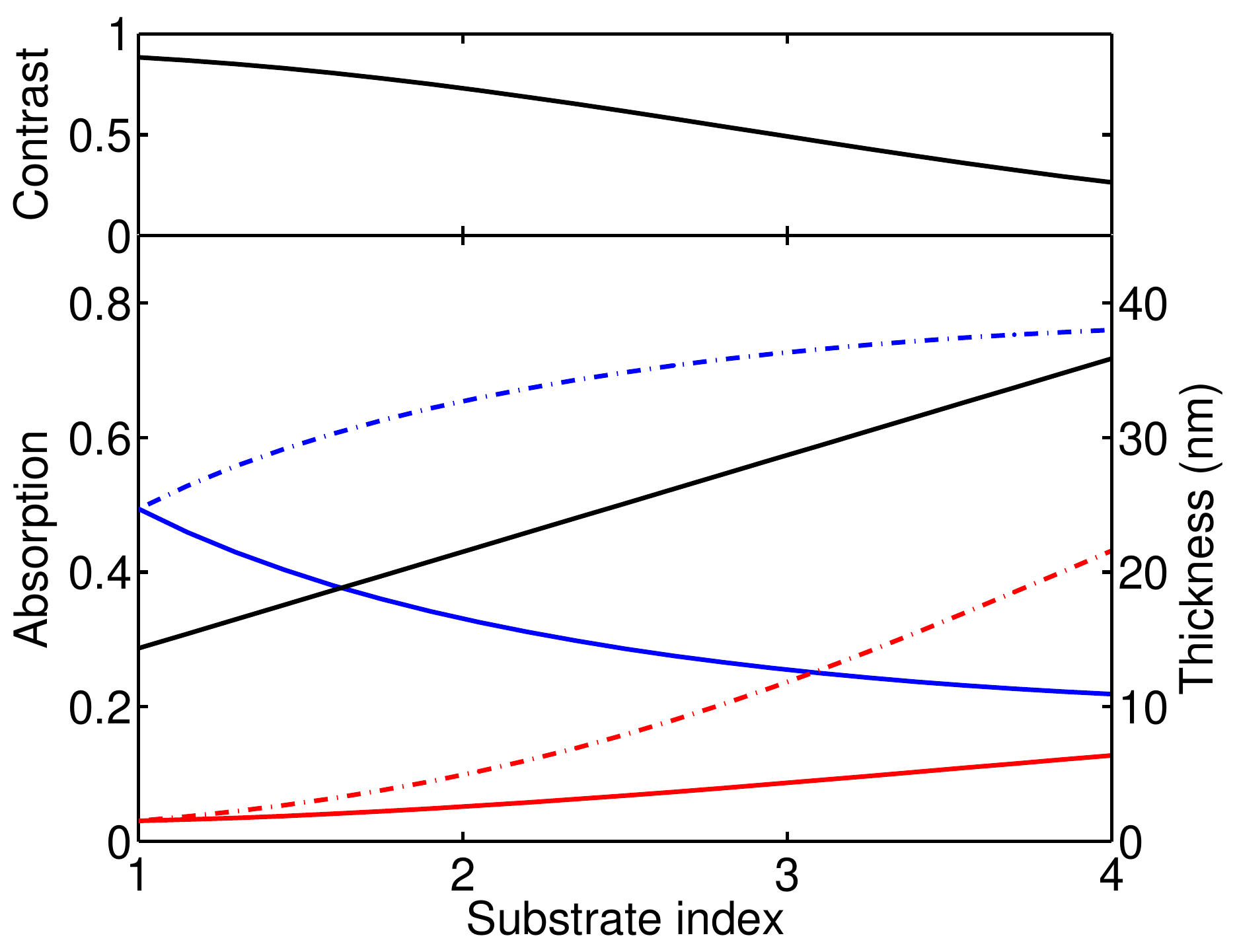}
\caption{Calculated absorption for a detector with filling factor
0.5, lattice period 200~nm, at a wavelength of 1550~nm, as a
function of the substrate refractive index. The solid curves are for
illumination from the air side, the dash-dotted curves for
illumination from the substrate side. The blue and red curves give
the absorption for polarization parallel and perpendicular to the
wires, respectively. The detector thickness is changed at each
substrate index, to achieve maximal absorption. The thickness is
given by the black line (right axis). The top graph shows the polarization contrast.\label{fig:substrate}}
\end{figure}

Commonly, NbN SSPDs are deposited on a substrate of sapphire and
illuminated from air. An inspection of Eq.~(\ref{eq:absorption})
shows that for a certain choice of sub- and superstrate, a factor
of $n_3/n_1$ in absorption can be gained by illuminating the
detector from the medium with the higher refractive index.
Figure~\ref{fig:substrate} shows the calculated absorption for a
detector, with a superstrate of air $(n_1=1)$, as a function of the
refractive index of the substrate. The thickness of the detector is
set such that
maximal absorption in the detector is achieved. This thickness is indicated with the black
line. The solid curves give the absorption for illumination from the
air, whereas the dash-dotted curves give the absorption for
illumination from the substrate. The blue and red curves are for
polarization parallel and perpendicular to the wires, respectively. We stress that this effect is caused by a lower impedance mismatch and should be separated from the cavity enhancement of the absorption, previously reported in Refs.~\cite{Rosfjord:2006p2886,Anant:2008p7675}. 

When the substrate index is increased, the absorption rises for
illumination from the substrate side. For illumination from the air side, the absorption for parallel polarization decreases. Note however that in both cases, the polarization contrast
decreases, from $C=0.88$ at $n_3=1$ to $C=0.26$ at $n_3=4$, and is independent on the direction of illumination, as shown in the top graph of Fig.~\ref{fig:substrate}. The absorption is a
factor of $n_3$ higher, when the detector is illuminated from the
substrate, as expected from the impedance model. It is interesting to note that, for parallel-polarized light, the absorptions from super-
and substrate add up to give $A_\mathrm{super}+A_\mathrm{sub}\approx
1$.
It is therefore possible to construct a detector with higher
absorption, up to 70\%, and lower polarization contrast, by using a high
refractive index substrate (e.g. Si or GaAs) and illuminating the
detector from the substrate. Unfortunately,
increasing the refractive index of the substrate also increases the
wavelength for which diffraction orders in the substrate appear. The first diffraction order at normal incidence appears at $\lambda/a = n_3$, with $a$ the periodicity of the structure, and $\lambda$ the wavelength of the light. In general,
these diffraction orders lower the absorption efficiency. For
a typical lattice period of 200~nm, and a substrate index of
$n_3=3.5$, the first diffraction order appears at a wavelength of
700~nm, making detectors on a high-refractive-index substrate less
effective for detecting visible light. The problem of diffraction could also be circumvented by designing a detector that has a variable line spacing.


\section{Conclusion}
In conclusion, we have measured a polarization dependence in the
detection efficiency of NbN superconducting single photon detectors
and find a wavelength dependent polarization contrast between 5\%
and 30\%. This effect can be explained by the geometry of the
detector. Calculations of the optical absorption efficiency  give good
agreement with the measured data. We have demonstrated that the
polarization dependence can be removed by the use of wedge
depolarizers. 

Furthermore, we have shown that the parameters of the detector can be
tuned to achieve an absorption for a polarization parallel to
the detector wires, that exceeds the absorption of an unpatterned
film of the same thickness. We have given a simple optical impedance model, that allows for
a quick estimate of the parameters needed to optimize the detector. For parallel-polarized light, the maximum absorption achievable is not determined by the
thickness or the dielectric constant of the metal film, nor by the
filling factor, but only by the refractive indices of the surrounding media. We have shown that by
illuminating the detector from the substrate it is possible to
increase the detection efficiency of the detector even further, by a
factor equal to the refractive index of the substrate. Such highly absorbing, highly polarization-dependent
detectors can be employed to efficiently detect photons with a well-defined polarization.

\section*{Acknowledgments}
We thank Jos Disselhorst, Jennifer Kitaygorsky, Teun Klapwijk, Hans
Mooij, and Raymond Schouten for technical assistance and
discussions. This research was financed by the Dutch Foundation for
Scientific Research (NWO) and the Foundation for Fundamental
Research of Matter (FOM).

\bibliography{nbn2}

\end{document}